\documentclass[%
 reprint,
superscriptaddress,
 amsmath,amssymb,
 aps,
pra,
]{revtex4-2}

\usepackage{orcidlink}
\usepackage{graphicx}
\usepackage{enumitem} 
\usepackage{dcolumn}
\usepackage{bm}
\usepackage{wasysym}
\usepackage{upgreek}
\usepackage{braket}

\usepackage[dvipsnames]{xcolor}
\usepackage[normalem]{ulem}

\usepackage{siunitx}
\usepackage{float}
\usepackage{xcolor}

\DeclareMathOperator{\Rb}{^{87}\text{Rb}}
\DeclareMathOperator{\Yb}{^{174}\text{Yb}}
\def\kB{k_\mathrm{B}}                    		        

\DeclareSIUnit\rad{rad}

\begin{document}

\preprint{APS/123-QED}

\title{Dual-species alkali and alkaline-earth-like optical tweezer arrays via interferometrically aligned high-NA objectives}

\author{K. Weber \orcidlink{0000-0003-2285-3840}}
\affiliation{Physics Department, University of Maryland, College Park, MD, 20742}
\affiliation{Joint Quantum Institute, National Institute of Standards and Technology, and University of Maryland, Gaithersburg, Maryland, 20899, USA}

\author{M. McMaster \orcidlink{0009-0004-3326-1124}}%
 \affiliation{Physics Department, University of Maryland, College Park, MD, 20742}
\affiliation{Joint Quantum Institute, National Institute of Standards and Technology, and University of Maryland, Gaithersburg, Maryland, 20899, USA}

\author{M. Anderson \orcidlink{0000-0001-7946-4227}}%
\altaffiliation[Present address: ]{IonQ, College Park, MD}
 \affiliation{Physics Department, University of Maryland, College Park, MD, 20742}
\affiliation{Joint Quantum Institute, National Institute of Standards and Technology, and University of Maryland, Gaithersburg, Maryland, 20899, USA}

\author{J. Tu~\orcidlink{0009-0006-4269-3230}}%
 \affiliation{Physics Department, University of Maryland, College Park, MD, 20742}
\affiliation{Joint Quantum Institute, National Institute of Standards and Technology, and University of Maryland, Gaithersburg, Maryland, 20899, USA}

\author{S. E. Eustice\ \orcidlink{0000-0003-1102-2400}}%
\affiliation{Joint Quantum Institute, National Institute of Standards and Technology, and University of Maryland, Gaithersburg, Maryland, 20899, USA}

\author{N.~A.~Schine}%
 \affiliation{Physics Department, University of Maryland, College Park, MD, 20742}
\affiliation{Joint Quantum Institute, National Institute of Standards and Technology, and University of Maryland, Gaithersburg, Maryland, 20899, USA}

\author{I.~B.~Spielman\ \orcidlink{0000-0003-1421-8652}}%
\affiliation{Joint Quantum Institute, National Institute of Standards and Technology, and University of Maryland, Gaithersburg, Maryland, 20899, USA}

\author{J.~V.~Porto}%
\affiliation{Joint Quantum Institute, National Institute of Standards and Technology, and University of Maryland, Gaithersburg, Maryland, 20899, USA}

\author{S.~L.~Rolston}%
 \affiliation{Physics Department, University of Maryland, College Park, MD, 20742}
\affiliation{Joint Quantum Institute, National Institute of Standards and Technology, and University of Maryland, Gaithersburg, Maryland, 20899, USA}

\author{{S.~Subhankar}\ \orcidlink{0000-0001-8664-3098}}
 \altaffiliation[Present address: ]{QuEra Computing, Boston, MA}
 \affiliation{Physics Department, University of Maryland, College Park, MD, 20742}
\affiliation{Joint Quantum Institute, National Institute of Standards and Technology, and University of Maryland, Gaithersburg, Maryland, 20899, USA}

\date{\today}

\begin{abstract}
We have developed a dual-species optical tweezer array apparatus combining $\Rb$ and $\Yb$ atoms with a permanent hybrid Twyman-Green--Fizeau interferometer, which provides precision co-alignment of two opposing 0.6-numerical aperture (NA) objectives and the high NA beams that pass through the system.
This technique is extensible to other tweezer platforms operating at high numerical aperture across widely-separated wavelengths.
Here we present simultaneous trapping and single-site resolved imaging of both species in co-aligned tweezer arrays with $\Rb$ confined at $840\ \rm{nm}$ and $\Yb$ at $532\ \rm{nm}$.
The platform provides a foundation for hybrid quantum register operation, including mid-circuit measurements and asymmetric intra- and inter-species interactions, greatly expanding the capabilities of neutral atom arrays.
\end{abstract}

\maketitle





\section{\label{sec:level1}Introduction}

Optical tweezer arrays of neutral atoms are one of the most versatile and rapidly advancing platforms for quantum simulation and quantum computation~\cite{grimm_optical_2000,weiss_quantum_2017,bluvstein_logical_2024,ebadi_quantum_2021,scholl_quantum_2021,nakamura_hybrid_2024,rozanov_benchmarking_2025,evered_highfidelity_2026,radnaev_universal_2024}, owing to their ability to arrange individual atoms into arbitrary geometries, perform single-site-resolved imaging, and execute high-fidelity entangling operations ~\cite{levine_highfidelity_2018,levine_parallel_2019,lester_rapid_2015}.
A frontier of growing importance is the extension of these capabilities to ``dual-species'' platforms, in which two co-located atomic species are simultaneously trapped, imaged, and manipulated.
Dual-species arrays offer new capabilities inaccessible to single-species systems, including species-selective operations, independent readout channels, and the ability to engineer interspecies interactions at the single-particle
level~\cite{anand_dualspecies_2024,beterov_rydberg_2015,brooks_preparation_2021,fang_interleaved_2025,liu_building_2018,nakamura_hybrid_2024,petrosyan_fast_2024,sheng_defectfree_2022,singh_dualelement_2022,spence_preparation_2022,wei_dualspecies_2024,wei_enhanced_2026,zhang_optical_2022}.

The combination of Rb and Yb is a well-motivated  dual-species pairing. As an alkali atom, Rb provides a mature toolbox: straightforward and economical laser cooling and optical trapping, accessible hyperfine qubits, and high-fidelity Rydberg-mediated two-qubit gates~\cite{bluvstein_publisher_2026,evered_highfidelity_2026,levine_highfidelity_2018,saffman_quantum_2010,browaeys_many-body_2020}. As an alkaline-earth-like atom, Yb offers complementary resources: a narrow intercombination transition at \qty{556}{\nm} suitable for laser cooling to $\simeq\unit{\upmu\kelvin}$ temperatures, a magnetic field-insensitive $\mathrm{^1S_0}$ ground and $\mathrm{^3P_0}$ metastable states which support highly coherent qubit operations ~\cite{jenkins_ytterbium_2022,saskin_narrowline_2019}.
A Rb-Yb platform leverages the strengths of both individual species and benefits from the wide tunability of the inter-species Rydberg interaction.
When combined with the predicted small intra-species $\mathrm{C_6}$ coefficients of the Yb singlet series Rydberg states, Yb can, for example, serve as an ideal mid-circuit readout qubit for Rb data qubits ~\cite{vaillant_long-range_2012}. Here we describe a dual-species optical-tweezer-array apparatus combining $\Rb$ and $\Yb$ atoms, with independently reconfigurable arrays of \qty{840}{\nm} tweezers and \qty{532}{\nm} tweezers respectively.

A $\Rb$ and $\Yb$ atom array comes with significant technical challenges. The components of the requisite optical systems must deliver micron-level alignment across the large wavelength range (\qty{308}{\nm} to \qty{1013}{\nm}) required for both species' cooling, trapping, imaging, and Rydberg excitation. Achieving this level of simultaneous alignment is particularly challenging. For example, it is difficult to align the two arrays when the species' imaging transitions are at widely disparate wavelengths. We solve these technical requirements with the permanent integration of a hybrid Twyman-Green--Fizeau interferometer, which we use to generate a precisely aligned reference beam normal to the surface of the glass cell (Fig~\ref{fig:Layout_fig}).
This enables the sequential alignment of two opposing 0.6 numerical aperture (NA) objectives and all of the tweezer projection and atom imaging optics. This process is entirely optics-based and requires no in-vacuum components or atomic signal.

We validate this approach by sequentially loading and imaging single $\Rb$ and $\Yb$ atoms simultaneously trapped in their respective tweezer arrays, achieving a detection fidelity of $99.97(3)\ \%$ for $\Rb$ and $99(1)$ for $\Yb$.
We determine the probability of surviving one image is $98.0(1.5)\ \%$ for Rb and $86(10)\ \%$ for Yb.

This manuscript is organized as follows.
First, Sec.~\ref{sec:apparatus} describes the apparatus including our novel laser cooling configuration. 
Then Sec.~\ref{sec:tweezers} continues by describing our experimental results on tweezer characterization and performance.
Finally, Sec.~\ref{sec:conclusion} summarizes and provides an outlook.

\section{Apparatus}\label{sec:apparatus} 
\begin{figure}[tb!]
    \centering
    \includegraphics{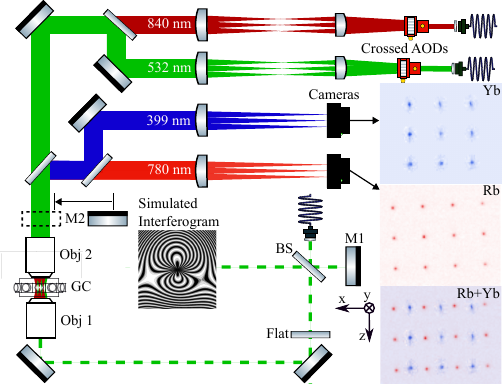}
    \caption{Optical system overview. 
    Laser light for the optical tweezers (maroon and green) is delivered via optical fiber, sent through independent AODs, and via Objective-2 (Obj-2), demagnified and sent to the vacuum system.
    Atomic florescence from both Rb and Yb (red and blue) is collected through Objective-2, separated by a dichroic beamsplitter, and independently imaged.
    The main TGF interferometer beam (green-dashed) travels in the opposite direction, is reflected by a insertable mirror (M2) before creating an interferogram, as pictured.
    }
    \label{fig:Layout_fig}
\end{figure}

The apparatus (detailed in~\cite{subhankar_engineering_2024}) is built around a fused-silica cell centered between two high-NA objectives, ``Objective-1'' and ``Objective-2'', as shown in Figure \ref{fig:Layout_fig}.
The cell consists of two parallel fused-silica nano-textured windows ($\diameter=\qty{82.5}{\mm}$, \qty{9.5}{\mm}~thick) that define the high-NA optical axis, and eleven smaller windows ($\diameter=\qty{18}{\mm}$), arranged in a dodecagonal geometry.
These provide low-NA optical access in the plane perpendicular to the optical axis.
The cell defines the geometry for our apparatus, such that the origin is at the geometric center of the cell, $\hat{z}$ is normal to the large windows, and $\hat{y}$ is along gravity.
The diameter of the large windows exceeds that of the objectives ($\diameter=\qty{50.8}{mm}$), providing off-axis optical access through the large windows, between the objective and the aperture of the large window. 
The thickness of the large windows balances two competing effects on aberrations: aberrations from window-objective mis-alignment increase with window thickness, while aberrations from vacuum-pressure-induced bowing of the windows decrease with window thickness.

Our cell windows are constructed with reactive-ion-etched nano-structured anti-reflection (RAR) glass~\cite{hobbs_contamination_2013,hobbs_continued_2012}.
The RAR-treated glass provides broad-band low reflectivity over the laser wavelengths required for our final system (\qty{308}{\nm} to \qty{1013}{\nm}). 
In addition, the RAR glass has low reflectivity even for large incident angles, which is particularly advantageous for high-NA optical paths and our shallow angle magneto-optical trap (MOT), see the \textit{Laser cooling} section below.

The $\mathrm{NA}=0.6$ objectives were optimized for a \qty{21.025}{mm} working distance including a \qty{2}{mm} air gap between the objective and the cell, and the \qty{9.5}{mm} thick fused-silica windows.
They were designed to provide diffraction-limited performance at six wavelengths required for high resolution trapping (\qty{850}{nm} and \qty{532}{nm}), and detection (\qty{420}{nm}/\qty{780}{nm} and \qty{399}{nm}/\qty{556}{nm}) of Rb and Yb, respectively. 

\vspace{3pt}\noindent
\textit{Interferometric Alignment}---Objective-2 projects the tweezer light into the cell, and collects atomic fluorescence for dual species imaging.
Objective-1 provides high-NA optical access for applications requiring enhanced fluorescence collection or additional control.  
In addition to necessitating accurate alignment to the cell, many two-objective applications require micron-level control of the relative positions of the objectives within their $\approx \qty{1}{\upmu\m}$ depth of focus along the optical axis, and  their $\approx$ \qty{200}{\upmu\m} $\times$ \qty{200}{\upmu\m} field of view along the transverse directions at multiple wavelengths.
Together these demand exceptional angular alignment.
We resolve the requirement of precision multi-wavelength alignment of the high-NA objectives using overlapping Twyman-Green and Fizeau (TGF) interferometers built into the apparatus (Fig.~\ref{fig:Layout_fig}). For more information on the interferometer design and its alignment, refer to \cite{subhankar_engineering_2024}.

Our hybrid TGF interferometer provides reference wavefronts that, along with a shear-plate interferometer, enables sequential high precision alignment of first the reference beam to the cell, and then the objectives to the reference beam. 
This approach achieves a residual angular deviation of $\lesssim 10\ \upmu \mathrm{rad}$ from normal incidence to the glass cell, greatly exceeding the $\lesssim 500\ \upmu \mathrm{rad}$ requirement for diffraction-limited performance.
The reference beam then serves as a common optical reference to which all remaining optical systems are independently aligned, significantly simplifying the construction of the apparatus and largely obviating the need to optimize the imaging system based on atomic signals.

\begin{figure}[tb!]
    \centering
    \includegraphics{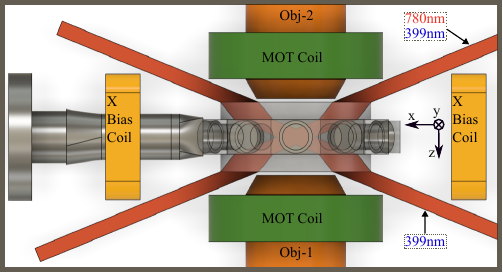}
    \caption{
    Apparatus side view including four shallow-angle MOT beams.
    The glass cell (translucent) is mounted via standard vacuum fittings (left).
    Shallow-angle MOT beams (red) narrowly avoid the objectives (orange) and MOT quadrupole coils (green) before entering through the top and bottom viewports.
    While 3D control of the magnetic field is provided by three pairs of bias coils, only the $x$-bias coils (yellow) are shown.
    Not shown are the out-of-plane MOT beams and through-objective beams at 532~nm, 556~nm and 840~nm beams.}
    \label{fig:MOT_Beam_Geometry}
\end{figure}

\vspace{3pt}\noindent
\textit{Laser cooling}---The large solid angle subtended by the objectives greatly constrains the configuration of the laser beams and magnetic-field-generating coils needed for the MOTs (see Fig.~\ref{fig:MOT_Beam_Geometry}).
The most natural, compact configuration of a standard 6-beam MOT nests the objectives inside the quadrupole coils, aligning the optical axis with the symmetry axis, $\hat{z}$, of the qudrupole fields. 
Since the high NA objective subtends much of the optical access along the axial direction, trapping and cooling along $\hat{z}$ in this geometry is usually provided by counter-propagating beams through the objectives~\cite{saskin_narrowline_2019,jenkins_ytterbium_2022}.
This aligns $\vec{B}$ and $\vec{k}$, which provides the most efficient trapping force using $100\ \%$ circularly polarized light.

In our case, the need to separate the scattered cooling light from the fluorescence imaging light for all three laser cooling wavelengths ($780\ {\rm nm}$, $556\ {\rm nm}$, $399\ {\rm nm}$) during imaging makes this challenging.
Instead, to trap and cool the atoms in 3D while avoiding sending all the MOT beams through the objective, we implemented a ``shallow-angle MOT" for $780\ {\rm nm}$ and $399\ {\rm nm}$, where the radial trapping is provided by standard orthogonal pairs of beams in the $x$-$y$ plane, but the axial trapping force is provided by beams entering the cell from outside of the objective, at an angle of $\theta_z=67.5$ degrees with respect to the optical axis. This angle is only $22.5$ degrees out of the $x$-$y$ symmetry plane, which leads to mixed, imperfect beam polarizations and significantly weaker axial confinement than the in-plane beams. We find that we can make a Rb MOT with only one pair of counter-propagating laser beams, but we  need two pairs of beams to trap Yb atoms on the $399\ {\rm nm}$ transition (see Fig.~\ref{fig:MOT_Beam_Geometry}).

Both atomic species are independently laser cooled in the tweezer region.
The shallow-angle $\Rb$ MOT is directly loaded from background vapor provided by dispensers.
Once loaded, these atoms undergo polarization gradient cooling (PGC) to $\approx\qty{30}{\upmu K}$. The shallow-angle ``blue'' $\Yb$ MOT is loaded from a 2D-MOT-generated cold atomic beam.
After loading, atoms are Doppler cooled using a normal-angle MOT on the narrow-line 556~nm transition, reaching temperatures of $\approx\qty{10}{\upmu K}$. We now describe the loading of these ultracold atoms into their respective optical tweezers. 

\section{Dual-species tweezer arrays}\label{sec:tweezers}

We create optical tweezer arrays using laser light at \qty{840}{\nm} for $\Rb$ and \qty{532}{\nm} for $\Yb$ on independent optical paths. 
Each beam passes through a pair of crossed acousto-optic deflectors (AODs) driven by independent arbitrary waveform generators. 
We expand the collimated beams to illuminate the objective's backplane, combined on a dichroic, and projected into the chamber by Objective-2 (Fig. \ref{fig:Layout_fig}).

The \qty{840}{\nm} tweezers are focused to a calculated $1/e^2$ radius of \qty{0.7}{\upmu\meter} with predicted trap depth of $\kB \times \qty{1}{m\kelvin}$, given the \qty{2}{\mW} per-tweezer optical power measured at input to the Objective-2.
In a similar way the \qty{532}{\nm} have a \qty{0.44}{\upmu\meter} calculated waist, and a depth of $\kB \times \qty{3}{m\kelvin}$ for \qty{15}{\mW} per tweezer.
The resulting dual-species optical tweezer arrays are interleaved and share a common interferometrically defined focal plane, ideally suited for simultaneous trapping, control, and imaging of both species.

Concurrently loading both species into the tweezers is challenging because light-assisted $\Rb$-$\Yb$ collisions present in co-located MOTs reduces the atom number of both.
In practice we load sequentially by first laser cooling and loading the $\Rb$ array, and repeating for $\Yb$, although the reverse produces the same results (see Appendix \ref{app:seq}).

We image tweezer-trapped single-$\Rb$ atoms by applying counter-propagating cross-polarized $780\ {\rm nm}$ fluorescence/1D-cooling beams with a \qty{0.5}{\mm} waist.
Objective-2 collects the resulting fluorescence; this objective is part of a $25\times$ magnification microscope focused on an electron-multiplying charge-coupled device (EMCCD) camera.
Similarly, we fluorescence-image individual $\Yb$ atoms using the broad-line blue transition with a pair of counter-propagating \qty{2.5}{\mm}-waist beams. 
Simultaneous cooling is provided by the green MOT beams; this partially mitigates recoil heating and reduces imaging-induced loss. 
Fluorescence is also collected by Objective-2 and imaged with $33\times$ magnification onto a second EMCCD camera.

\begin{figure}[t]
    \centering
    \includegraphics[]{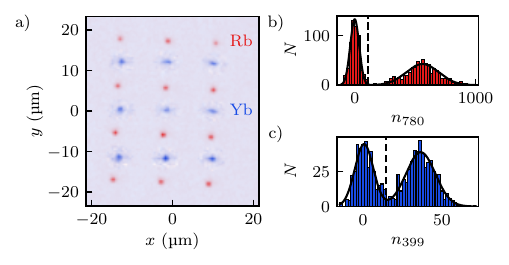}
    \caption{
    Simultaneous trapping and single-site-resolved imaging of
    $\Rb$ and $\Yb$ in co-aligned $4\times3$ and $3\times3$ tweezer arrays.
    (a) Composite fluorescence image. 
    We average images of simultaneously loaded $\Rb$ and $\Yb$ arrays over 100 repetitions of the experiment, and combine them into the displayed two-channel color image.
    (b) and (c) Histograms demonstrating sub-Poissonian loading statistics for $\Rb$ and $\Yb$ respectively.
    These histograms are compiled for the whole tweezer array, and are broadened from spatial inhomogeneities.
    In both cases, the horizontal axis plots the photon counts per trap.
    The solid lines shows a fit of the histogram to two displaced Gaussian distributions, while the dashed vertical lines shows the average detection threshold.
    }
    \label{fig:dual-species-twz-fig}
\end{figure}

Figure~\ref{fig:dual-species-twz-fig}(a) shows a composite fluorescence image of co-aligned single-atom arrays ($4\times3$ for $\Rb$ and $3\times3$ for $\Yb$) averaged over 100 experimental repetitions.
We verified the alignment of the tweezer planes by transferring Yb atoms from initial $532\ {\rm nm}$ traps, into $840\ {\rm nm}$ traps, and back.
We achieved a two-way transfer efficiency of up to 85(10)\ \%. 
Displacing the arrays by just $\approx\qty{1}{\um}$ transversely or $\approx\qty{1.5}{\um}$ longitudinally reduces the two-way transfer efficiency by more than half and doubling those displacements leads to zero transfer efficiency.
This demonstrates the capability of our interferometer for 3D alignment of independent arrays at a distance scale suitable for Rydberg-based gates. Under these conditions, we observe array-average loading rates of both $\Rb$ and $\Yb$ of $53(5)\ \%$ and $57(7)\ \%$ respectively.

The aggregate photon count distributions in the neighborhood of the tweezer traps exhibit two well-resolved peaks corresponding to zero and one trapped atom for both species (Fig.~\ref{fig:dual-species-twz-fig}).
The similar contributions of each peak stem from the roughly equal loading probability of zero and one atom in the collisional-blockade
regime for tightly focused tweezers~\cite{schlosser_subpoissonian_2001,schlosser_collisional_2002}. 
For $\Rb$ the overlap integral of the fitted distributions yields an array-averaged, loss-excluded, discrimination fidelity of $99.97(3)\ \%$, whereas for $\Yb$ we measure a detection fidelity of $99(1)\ \%$.
For further details on the image processing and the determination of imaging infidelities, see appendix~\ref{sec:image-processing}.

For both $\Rb$ and $\Yb$ laser cooling is operative during imaging (1D PGC for $\Rb$, and 3D Doppler cooling on the 556~nm transition for $\Yb$), mitigating recoil driven heating and loss.
For Rb and Yb, we find single-image survival probabilities of $98.0(1.5)\ \%$ and $86(10)\ \%$ respectively, by determining the fraction of atoms present in back-to-back images on a per-tweezer basis.
These survival probabilities have not been fully optimized for this initial report.

This demonstrates simultaneous trapping and single-site-resolved imaging of an alkali and an alkaline-earth-like atom in co-aligned optical tweezer arrays, establishing this platform as a foundation for experiments exploiting independent control and measurement as well as asymmetric inter- and intra-species interactions~\cite{vaillant_long-range_2012}.

\section{Conclusion and outlook}\label{sec:conclusion}

We have presented a dual-species optical-tweezer apparatus combining $\Rb$ and $\Yb$.
The built-in Twyman-Green--Fizeau interferometer provides a common optical reference for aligning the two opposing 0.6~NA objectives without requiring an atomic signal and permits in situ recalibration.
Using the interferometer, we align the tweezer beams normal to the cell faces and co-locate the objective focal planes.
Our combination of interferometric dual-objective alignment and species-selective cooling, trapping, and imaging provides a general approach that is applicable beyond the $\Rb$--$\Yb$ combination.

Our specific platform enables a wide range of experiments.
For example, the $\Rb$--$\Yb$ pair provides a candidate platform for hybrid quantum registers in which the two species play complementary roles.
Rubidium atoms offer a mature toolbox for Rydberg-mediated entanglement and high-fidelity single-qubit control, making them natural candidates for computational qubits.
By contrast, ytterbium provides an optical clock transition and a zero-electronic-angular-momentum ground state, offering complementary resources for optical control and metrology.
The large difference in transition frequencies between $\Rb$ and $\Yb$ supports species-selective readout and optical control of both species.

Our addition of a second high-NA objective supports future applications including enhanced fluorescence collection and external half-cavity formation, atom--photon entanglement generation, species-selective ac Stark addressing, and cavity quantum electrodynamics~\cite{soper_stability_2026,faorlin_controlling_2025,slodicka_atomatom_2013}.

\begin{acknowledgments}
The authors thank A.~Restelli for extensive technical assistance, and T.-C.~Tsui for initial construction work.
This work was partially supported by the National Institute of Standards and Technology; the National Science Foundation through the Quantum Leap Challenge Institute for Robust Quantum Simulation (grant OMA-2120757); and the Air Force Office of Scientific Research Multidisciplinary University Research Initiative ``RAPSYDY in Q'' (No. FA9550-22-1-0339).
\end{acknowledgments}

\bibliography{main}

\appendix

\setcounter{equation}{0}
\setcounter{figure}{0}
\setcounter{table}{0}

\renewcommand\theequation{A\arabic{equation}}
\renewcommand\thefigure{A\arabic{figure}}

\section{Tweezer Loading Sequences}
\label{app:seq}

In this section, we expand on the main text's discussion on the preparation and loading of $\Rb$ and $\Yb$ atoms: starting with the initial laser cooling and ending with tweezer-trapped atoms ready for science. 

\vspace{3pt}\noindent
\textit{Rubidium}---We introduce Rb into the system using RB/NF/4.8/17 dispensers from SAES Getters.
The $\Rb$ MOT is loaded directly from the resulting vapor in $3\ {\rm s}$ using $\approx 780\ {\rm nm}$ light on the ${\rm5s}\ ^2{\rm S}_{1/2}\rightarrow\, {\rm5p}\ ^2{\rm P}_{3/2}$, i.e., D2, transition.
The MOT beams are in the shallow-angle configuration described in Sec.~\ref{sec:apparatus}, and
all contain both cooling and repumping light on the $\ket{F=2}\rightarrow\ket{F'=3}$ and $\ket{F=1}\rightarrow\ket{F'=2}$ transitions respectively.
The cooling light is red-detuned $\Delta\approx10\Gamma$ with $I/I_\text{sat}\approx 14$, where $\Gamma = 36\times10^6\ {\rm s}^{-1}$ is the $5{\rm P}_{3/2}$ decay rate, and $I_\text{sat} \approx 1.7\ {\rm mW}/{\rm cm}^2$ is the saturation intensity.
The MOT uses a $12\ {\rm G}/{\rm cm}$ magnetic field gradient ($10\ {\rm G} = 1\ {\rm mT}$).

We prepare for tweezer loading by linearly ramping the magnetic field gradient to zero in $6\,{\rm ms}$ and reducing the saturation parameter to $\approx7$.
During the following $3\ {\rm ms}$ of sub-Doppler cooling (i.e., PGC) the atoms' temperature falls to $\approx30\,\upmu\text{K}$.
Simultaneously, the array of $2\,\text{mK}$-deep 840 nm tweezer traps is switched on and loaded.
After this time, the $780\ {\rm nm}$ light is extinguished, allowing the untrapped atoms to fall away for $30\ {\rm ms}$.

The trapped $\Rb$ atoms are imaged for $50\ {\rm ms}$ with a pair of $0.5$ mm left-circularly polarized counter-propagating beams containing both cooling and repumping light.

\vspace{3pt}\noindent
\textit{Ytterbium}---We generate a wide-angle Yb atomic beam with a $390\ {\rm C}$ oven, and then capture the low-energy tail using a blue 2D MOT on the $399\ {\rm nm}$ $\rm{6s^2}\ ^1\rm{S}_0\rightarrow\,\rm{6s6p}\ ^1\rm{P}_1$ transition.
Permanent magnets create the magnetic field gradient.
A 399 nm push beam directs a cold beam from the 2D MOT into the science cell where they are captured in a hybrid $399\ {\rm nm}$ / $556\ {\rm nm}$ 3D MOT.
The latter addresses the narrow green $\rm{6s^2}\ ^1\rm{S}_0\rightarrow\,\rm{6s6p}\ ^3\rm{P}_1$ transition; on this transition, the shallow-angle configuration has anti-trapping forces that are too great to overcome, as a result we fall back to a conventional MOT geometry, with normal-incidence beams counter-propagating through the objectives (currently, this prevents us from green imaging of $\Yb$).
This MOT loading stage takes just $0.5\ \rm{s}$.

The $399\ {\rm nm}$ light is $\approx7\Gamma$ red-detuned with $I/I_{\text{sat}}\approx 0.5$ (with $\Gamma = 192\times10^6\ \rm{s}^{-1}$ and $I_\text{sat}=57\,\rm{mW}/{cm}^2$ for the $\rm{6s^2}\ ^1\rm{S}_0\rightarrow\,\rm{6s6p}\ ^1\rm{P}_1$ transition) and the 556 nm light is around $3\Gamma$ red-detuned and much more saturated, with $I/I_{\text{sat}}\simeq 500$ (with $\Gamma = 1.15\times10^6\ \rm{s}^{-1}$ and $I_\text{sat}=0.14\ \rm{mW}/\rm{cm}^2$ for the $\rm{6s^2}\ ^1\rm{S}_0\rightarrow\,\rm{6s6p}\ ^3\rm{P}_1$ transition).
The $\Yb$ MOT uses a magnetic field gradient of $26\ {\rm G}/{\rm cm}$, and the final MOT is offset by about $200\ \upmu{\rm m}$ from the tweezer region.

We then transfer to a green MOT in $4\ {\rm ms}$ by turning off the $399\ {\rm nm}$ beams with an exponential ramp, linearly reducing the magnetic field gradient to $7.5\ {\rm G}/{\rm cm}$, and linearly ramping the $556\ {\rm nm}$ detuning and intensity to $\approx -1.7\Gamma$ and $\approx 12 I_{\text{sat}}$, respectively.
Because the $\Yb$ MOTs is displaced from the tweezer region it must be moved to overlap with the array; we do with a $40\ {\rm ms}$ ramp of the magnetic fields along with a linear ramp of the detuning and intensity to $\approx -1.6\Gamma$ and $\approx 0.25\,I_{\text{sat}}$, respectively.

The final MOT temperature is $\approx8\upmu\text{K}$, enabling direct loading of our $2\ {\rm mK}$ deep $532\ {\rm nm}$ tweezer array in $30\ {\rm ms}$. 
The MOT gradient and cooling light are then abruptly turned off, and we allow $30\ {\rm ms}$ for the untrapped atoms to fall away.
Lastly, a $10\ {\rm ms}$ blow-away pulse using $556\ {\rm nm}$ light at low intensity ($0.18\,I_{\text{sat}}$) stimulates light-assisted collisions, ensuring an occupancy of at most one atom per trap.

For imaging, the atoms are illuminated with a pair of counter-propagating $399\ {\rm nm}$ beams at ultra-low intensity ($0.005 \,I_{\text{sat}}$ each) red detuned by $7\ \Gamma$ from resonance for $250\ \rm{ms}$.
We turn the $556\ {\rm nm}$ MOT beams on during imaging for simultaneous cooling ($I = 0.18 \,I_{\text{sat}}$, and $\Delta = -1.7\ \Gamma$). 
Using a long exposure with a low-intensity far detuned blue beam yields a higher survival probability than a brief high intensity resonant imaging pulse.
However, rapid $399\ {\rm nm}$ has been demonstrated by other groups~\cite{muzifalconi_microsecondscale_2025, yokoyama_minimally_2026}, providing a clear path for improvement.

\begin{figure*}
    \centering
    \includegraphics[]{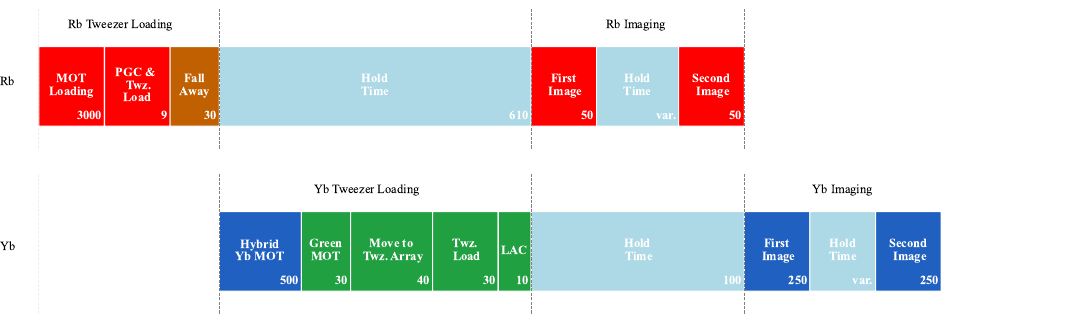}
    \caption{Representative experimental sequence for simultaneous loading, with step durations listed in ms. Alternative experiments have interlaced the imaging of $\Rb$ and $\Yb$ with no effect on atomic detection, loading or survival.}
    \label{fig:seq}
\end{figure*}

\vspace{3pt}\noindent
\textit{Simultaneous loading}---
As seen in Fig.~\ref{fig:seq}, the current process of dual species tweezer-loading is sequential: the $\Rb$ tweezers are fully loaded before $\Yb$ MOT loading even begins (the reverse sequence has been demonstrated as well). This choice is made primarily to avoid keeping $\Yb$ in tweezers for seconds while $\Rb$ loads. There is a trade-off in this choice, as it requires the $\Rb$ temperature to be very low and the $840\ \rm{nm}$ traps to be sufficiently deep to minimize loss during $\Yb$ loading.

The choice to load the MOTs sequentially is informed by preliminary experiments in our device demonstrate that both the $\Yb$ blue and green MOTs degrade the $\Rb$ MOT. This confirms analogous alkali-alkaline earth-like results from other groups \cite{nemitz_production_2009, witkowski_photoionization_2018, aoki_photoionization_2013}. For the blue MOT, this phenomenon is well-understood to be due to the energy associated with the $399\ \rm{nm}$ exceeding the photoionization threshold of $\Rb$ atoms in the $5^2\rm{P}_{3/2}$ state ($479\ \rm{nm}$) \cite{aoki_photoionization_2013}, reducing the atom number in the $\Rb$ MOT. In some cases, this could lead to losses in traps of both species as molecules form \cite{nemitz_production_2009}.

This hinders the simultaneous loading of spatially co-located MOTs. This in addition to both species' MOTs requiring significantly different magnetic field gradients to form optimally made sequential loading seem preferable to sequences requiring simultaneous MOT loading.

\section{Image processing}
\label{sec:image-processing}

In this section, we describe the analysis of camera data that is used to generate the images of the atom arrays. For each realization of the array, three images are taken for each species using the experimental protocol described in the previous section. The first two images contain the atoms and serve to measure the loading into the array and survival through the imaging process, while the third image serves as a background image. This background image is taken after briefly turning off both atom arrays and allowing the atoms to fall out of the trapping region before turning on the imaging beams. The electron-multiplying gain of each camera was calibrated by taking images under the same illumination conditions while varying the set gain value to determine how many additional electron-multiplied electrons were generated by the gain process.

Atom locations are automatically detected by taking the average of the first set tweezer array images, applying a Gaussian smoothening to this average, and then using a local maximum filter to determine an initial guess of the location of each tweezer in the image as well as the size of the point spread function of the imaging system. The image is then least-squares fit to a set of $N$ 2d Gaussian peaks corresponding to the $N$ tweezers, with the previously identified locations and peak amplitudes serving as the initial guess for a least-squares minimizer. All of the Gaussian peaks share a common standard deviation in $x$ and $y$ but have variable positions and peak heights. The resulting fits are used to construct weighted masks for each tweezer, with each pixel being weighted by a peak-normalized Gaussian distribution, reducing the amount of background photons that are counted into our signals.

Our dual species atom images are composites from two EMCCD cameras; we register the two camera planes to each other using a tweezer beam monitoring camera that samples tweezer light directly ahead of the objective. We have verified that this alignment is robust through to the atom plane to within a micron by transferring $\Yb$ atoms back and forth between the 532\ nm traps and the 840\ nm traps.

By summing up the weighted photon number for each tweezer in each image, a per-tweezer photon count is established. These photon counts are binned into histograms on a tweezer-by-tweezer basis, and each of these histograms are fit to the sum of two Gaussians. We determine a threshold by minimizing the sum of the false positive and false negative infidelity, which are defined as the area of the lower histogram above the threshold and the area of the upper histogram below the threshold, respectively. We note that these thresholds vary between 55 and 140 across the $\Rb$ array and between 4 and 8 across the $\Yb$ array, justifying doing this optimization on a mask-by-mask basis, rather than array-wide. Given these thresholds, we determine the presence or absence of an atom from data by direct comparison of the counts detected in a single mask on an individual run of the experiment with the threshold for that mask location. Survival fraction is defined as the total number of atoms detected in the second images of a dataset divided by the total number of atoms detected in the first image of the same dataset.

Having performed the above analysis on the raw images, background subtraction is done for the two arrays. For $\Rb$, a principal component analysis (PCA) basis is constructed from the background images. After removing an extended region of interest (ROI) around each tweezer location (which is four standard deviations of the Gaussian fit large) from both the PCA basis vectors and the atom array images, least squares fitting is done to determine the best PCA background image corresponding to each $\Rb$ image. For $\Yb$, the background images contain significantly different spatial structure than the atom array images, so a PCA analysis cannot be completed. For these images, an average no-atom background is constructed, first by averaging all of the images outside of the extended tweezer ROIs. For constructing a background image inside of each tweezer ROI, the aforementioned detection threshold is used to select images where no atom is present in a given tweezer. By averaging the zero-atom images for each tweezer, a full average background image is constructed for $\Yb$. This procedure of background reconstruction has a systematic error, where atoms lost early on in the imaging pulse could fall below the detection threshold but incorrectly inflate the number of photons that are added into the background image. We leave correction of this effect to future work. For both $\Rb$ and $\Yb$, the corresponding background image is then subtracted off before proceeding to photon counting analysis.

The process of fitting the average peaks, counting photons per tweezer, fitting photon count histograms and determining the infidelity, loading, and survival rate for each tweezer is then repeated on the background-subtracted images, generating the data presented in this paper.

\end{document}